\documentstyle[aps,preprint]{revtex}
\begin{document}
\draft

\title{Solution generating with perfect fluids }

\author{David Garfinkle}
\address{
\centerline{Department of Physics, Oakland University,
Rochester, MI 48309}}

\author{E.N. Glass}
\address{\centerline{Department of Physics, University of Michigan, 
Ann Arbor, MI 48109}
\centerline{and Physics Department, University of Windsor, 
Windsor, Ontario, N9B 3P4 (permanent address)}
}

\author{J.P. Krisch}
\address{
\centerline{Department of Physics, University of Michigan, Ann Arbor, MI 48109}
}
\maketitle

\null\vspace{-1.75mm}

\begin{abstract}
We apply a technique, due to Stephani, for generating solutions of the Einstein-perfect-fluid
equations.  This technique is similar to the vacuum solution generating techniques of Ehlers, Harrison, Geroch and others.  We start with a ``seed''
solution of the Einstein-perfect-fluid equations with a Killing vector.
The seed solution must either have (i) a spacelike Killing vector and
equation of state
$ P = \rho $
or (ii) a timelike Killing vector and equation of state
$ \rho + 3 P = 0 $.
The new solution generated by this technique then has the same Killing vector
and the same equation of state.  We choose several simple seed solutions
with these equations of state and where the Killing vector has no twist.  
The new solutions are twisting versions of the seed solutions.
\end{abstract}

\section{Introduction}

In solving Einstein's equation it is usual to treat spacetimes that have
one or more Killing vectors.  The existence of a Killing vector reduces 
Einstein's equation from a 4 dimensional set of equations to a 3 dimensional
set.  Instead of the spacetime metric one considers an equivalent set of
variables: the norm
$ \lambda $
and twist potential
$ \omega $
of the Killing vector and a metric
$ \gamma _{ab} $
on the 3 dimensional manifold of trajectories of the Killing field.  The vacuum
Einstein equations then become a set of equations for the quantities
$ \left ( \lambda , \, \omega , \, {\gamma _{ab}} \right ) $.
Remarkably, as shown by Ehlers\cite{ehlers} , Harrison\cite{harrison} , 
Geroch\cite{geroch} and others\cite{kramer1} 
these equations have a symmetry that allows us to 
generate new solutions: given
$ \left ( \lambda , \, \omega , \, {\gamma _{ab}} \right ) $
solving the equations then
$ \left ( {\tilde \lambda } , \, {\tilde \omega }, \, {\gamma _{ab}} \right ) $
is also a solution where
$ \left ( {\tilde \lambda } , \, {\tilde \omega } \right ) $ 
is a simple algebraic function of 
$ \left ( \lambda , \, \omega \right ) $.
These results have been generalized to the Einstein-Maxwell 
equations\cite{harrison,hauern,kramer2} and have been 
applied to find many new solutions.  (The general invariance
transformation was first published in \cite{kramer1} (in the vacuum case
with two Killing vectors) and in\cite{kramer2} (in the Einstein-Maxwell
case with one Killing vector).  Ehlers\cite{ehlers} found a special case
of the general vacuum transformation while Harrison\cite{harrison} found
a special case of the general Einstein-Maxwell transformation.) 

The vacuum solution generating technique was generalized to the case of
perfect fluids by Stephani\cite{stephani}.  This technique is very 
similar to that for
the vacuum Einstein equation.  However, it turns out that not all equations of
state are suitable for this sort of solution generating.  Only two equations of
state are compatible with this technique: 
$ P = \rho $
for a spacelike Killing field and 
$ \rho \, + \, 3 P = 0 $
for a timelike Killing field.

In section 2 we define our notation.  Sections 3 and 4 contain the new 
solutions that we have generated using this technique with 
section 3 treating the 
$ P = \rho $
case and section 4 treating the
$ \rho \, + \, 3 P = 0 $
case.  Section 5 contains a brief discussion of our results.

\section{Formalism}

Our notation is essentially that of refs.\cite{geroch,stephani}.  We use
signature ($-,+,+,+$) and units where $ c = G = 1 $.  Let
($M, g_{ab} $)
be a solution of the Einstein-perfect fluid equations with energy density
$ \rho $
and pressure
$ P $.
Assume that 
$ g_{ab} $
has a Killing vector
$ \xi ^a $.
Define the norm 
$ \lambda $
and twist
$ \omega _a $
of 
$ \xi ^a $
by
$ \lambda := {\xi ^a} {\xi _a} $
and
$$
{\omega _a} := {\epsilon _{abcd}} {\xi ^b} {\nabla ^c} {\xi ^d} \; \; \; .
\eqno(2.1)
$$
We assume that either (i) the Killing vector is orthogonal to the fluid four-velocity and 
$ P = \rho $
or (ii) the Killing vector is parallel to the fluid four-velocity and 
$ \rho + 3 P = 0 $.
From this assumption it follows that 
$ {R_{ab}} {\xi ^b} = 0 $.
It then follows that there is a scalar
$ \omega $
such that
$ {\omega _a} = {\nabla _a} \omega $
and that there are forms
$ \alpha _a $
and
$ \beta _a $
satisfying
$$
{\nabla _{[a}} {\alpha _{b]}} = {1 \over 2} \; {\epsilon _{abcd}} \, {\nabla ^c} {\xi ^d} \; \; \; ,
\eqno(2.2)
$$
$$
{\xi ^a} {\alpha _a} = \omega \; \; \; ,
\eqno(2.3)
$$
$$
{\nabla _{[a}} {\beta _{b]}} = 2 \, \lambda \, {\nabla _a} {\xi _b} \; +  \; \omega \, {\epsilon _{abcd}} \, {\nabla ^c} {\xi ^d} \; \; \; ,
\eqno(2.4)
$$
$$
{\xi ^a} {\beta _a} = {\omega ^2} \, + \, {\lambda ^2} \, - \, 1 \; \; \; .
\eqno(2.5)
$$

For any constant $ \theta $
define $ \tilde \lambda $
and 
$ \eta _a $
by
$$
\lambda / {\tilde \lambda } := {\cos ^2} \theta \, + \, \left ( {\omega ^2} + {\lambda ^2} \right ) {\sin ^2} \theta \, - \, 2 \omega \sin \theta \cos \theta \; \; \; ,
\eqno(2.6)
$$
$$
{\eta _a} := {{\tilde \lambda }^{ - 1}} \, {\xi _a} \; + \; 2 \, \cos \theta \, \sin \theta \, {\alpha _a} \; - \; {\sin ^2} \theta \, {\beta _a} \; \; \; .
\eqno(2.7)
$$
Then the new metric is given by
$$
{{\tilde g}_{ab}} = {\lambda \over {\tilde \lambda }} \; \left ( {g_{ab}} \; - \; {\lambda ^{ - 1}} \, {\xi _a} {\xi _b} \right ) \; + \; {\tilde \lambda } \, {\eta _a} {\eta _b} \; \; \; .
\eqno(2.8)
$$
This new metric is also a solution of the Einstein-perfect fluid equations
with the same equation of state and the same Killing vector.

One of the uses of solution generating techniques is that they can produce
twisting solutions from seed solutions that are twist free.  We now consider
how the process for finding the new metric simplifies in the case where the
seed metric is twist free.  In this case
$ \omega = 0 $
and 
$ {\nabla _a} {\xi _b} = {\lambda ^{ - 1}} \, {\xi _{[b}} {\nabla _{a]}} \lambda $.
Define 
$ F := {\cos ^2} \theta \, + \, {\lambda ^2} \, {\sin ^2} \theta $.
Then 
$ {\tilde \lambda } = \lambda / F $.
The equation for 
$ \beta _a $
can be solved as
$ {\beta _a} = \left ( \lambda \, - \, {\lambda ^{ - 1}} \right ) {\xi _a} $.
The equation for 
$ \alpha _a $
simplifies to 
$$
{\nabla _{[a}} {\alpha _{b]}} = {1 \over {2 \lambda }} \; {\epsilon _{abcd}} \, {\xi ^d} \, {\nabla ^c} \lambda
\eqno(2.9)
$$
and
$ {\xi ^a} {\alpha _a} = 0 $. The one-form
$ \eta _a $
is then given by
$$
{\eta _a} = {\lambda ^{ - 1}} \, {\xi _a} \; + \; 2 \, \cos \theta \, \sin \theta \, {\alpha _a}  \; \; \; .
\eqno(2.10)
$$
The new metric
is then given by 
$$
{{\tilde g}_{ab}} = F \; \left ( {g_{ab}} \; - \; {\lambda ^{ - 1}} \, {\xi _a} {\xi _b} \right ) \; + \; {\lambda \over F} \; {\eta _a} {\eta _b} \; \; \; .
\eqno(2.11)
$$
From now on we will assume that the seed solution is twist-free.

\section{$ P = \rho $ fluids}

\subsection{Fluid variables}

In describing the new solutions we will concentrate on two properties of 
interest: singularities and fluid flow.  The solution generating procedure
often results in singularities on the axis.  For each of our solutions we
will consider whether such a singularity exists.  A perfect fluid 
is described by the energy density $ \rho $, pressure $ P $ and fluid
four velocity $ u^a$.  The covariant derivative of the fluid four velocity is decomposed\cite{exact} as
$$
{\nabla _b} {u_a} = - \, {a_a} {u_b} \; + \; {\omega _{ab}} \; + \; {\sigma _{ab}} \; + \; \Theta \left ( {g_{ab}} \, + \, {u_a} {u_b} \right ) /3 \; \; \; .
\eqno(3.1)
$$
Here 
$ {a_a} := {u^b} {\nabla _b} {u_a} $
is called the acceleration.  The tensors
$ {\sigma _{ab}} $ (called the shear) and
$ \omega _{ab} $ (called the vorticity) are orthogonal to $ u^a $
with $ \omega _{ab} $ antisymmetric and $ \sigma _{ab} $
symmetric and trace-free.  The scalar $ \Theta $
is called the expansion.  Under solution generating with the
$ P = \rho $ equation of state the fluid parameters transform as follows:
$$
{\tilde \rho} = \rho / F \; \; \; ,
\eqno(3.2)
$$
$$
{{\tilde u}_a} = {\sqrt F} \, {u_a} \; \; \; ,
\eqno(3.3)
$$
$$
{\tilde \Theta } = {1 \over {\sqrt F}} \; \left ( \Theta \; + \; {F^{ - 1}} \, \lambda \, {\sin ^2} \theta \, {u^a} {\nabla _a} \lambda \right ) \; \; \; ,
\eqno(3.4)
$$
$$
{{\tilde a}_a} =  \, {a_a} \; + \; {F^{ - 1}} \, \lambda \, {\sin ^2} \theta \, 
\left ( {\nabla _a} \lambda \, + \, {u_a} \, {u^b} {\nabla _b} \lambda \right ) \; \; \; ,
\eqno(3.5)
$$
$$
{{\tilde \omega }_{ab}} = {\sqrt F} \, {\omega _{ab}} \; \; \; .
\eqno(3.6)
$$
There does not seem to be a simple transformation rule for
$ \sigma _{ab} $. 

A fluid with the equation of state $ P = \rho $
is called ``stiff matter.''  In such a fluid the adiabatic sound speed is equal
to the speed of light.  If the fluid is irrotational then the field equations are the same as those of a minimally coupled scalar field with timelike gradient.\cite{taub}  There is a technique for generating stiff matter 
solutions from vacuum solutions.\cite{wainwright}

\subsection{Robertson-Walker seed metric}

Consider the seed metric
$$
d {s^2} = - \; d {t^2} \; + \; {t^{2/3}} \; \left [ d {z^2} \; + \; d {r^2} \; + \; {r^2} \, d {\phi ^2} \right ] \; \; \; .
\eqno(3.7)
$$
This is a flat Friedmann-Robertson-Walker metric with
$ P = \rho $.
The fluid four velocity is 
$ {u_a} = - \, {\nabla _a} t $
and the energy density is
$ \rho = 1/\left ( 24 \pi {t^2} \right ) $.
The shear, vorticity and acceleration all vanish.  The expansion is
$ \Theta = 1/t $.
The metric is type O.
This metric has more than one Killing vector.  We first consider the case
$ {\xi ^a} = {{(\partial / \partial z )}^a} $.
Then we have
$ \lambda = {t^{2/3}} $
and
$ \omega = 0 $.
It then follows that
$ F = {\cos ^2} \theta \; + \; {t^{4/3}} \,  {\sin ^2} \theta  $.
The equation for 
$ \alpha _a $
is then
$$
{\nabla _{[a}} {\alpha _{b]}} = - \; {2 \over 3} \; r \, {\nabla _{[a}} r \, {\nabla _{b]}} \phi 
\eqno(3.8)
$$
(as well as $ {\xi ^a} {\alpha _a} = 0 $).
A solution is 
$$
{\alpha _a} = - \; {1 \over 3} \; {r^2} \; {\nabla _a} \phi \; \; \; .
\eqno(3.9)
$$
The one-form
$ \eta _a $
is then given by
$$
{\eta _a} = {\nabla _a} z \; - \; {2 \over 3} \; \cos \theta \, \sin \theta \, {r^2} \, {\nabla _a} \phi \; \; \; .
\eqno(3.10)
$$
The new metric 
$ {\tilde g}_{ab} $
is then given by
$$
d {{\tilde s}^2} = F \; \left [ - \; d {t^2} \; + \; {t^{2/3}} \; \left ( d {r^2} \; + \; {r^2} \, d {\phi ^2} \right ) \right ] \; + \;  {{t^{2/3}} \over F} \; {{\left ( d z \; - \; {2 \over 3} \; \cos \theta \, \sin \theta \, {r^2} \, d \phi \right ) }^2} \; \; \; .
\eqno(3.11)
$$
This new metric has no singularities on the axis.  The acceleration and vorticity vanish.  The expansion is 
$$
{\tilde \Theta} = {1 \over {\sqrt F}} \; \left ( {t^{ -1}} \; + \; {2 \over 3} \; {F^{- 1}} {t^{1/3}} {\sin ^2} \theta \right ) \; \; \; .
\eqno(3.12)
$$
The shear has components in the $r r , \, \phi \phi , \, zz $
and $z \phi $ directions.  Its square is
$$
{{\tilde \sigma}^{ab}} {{\tilde \sigma }_{ab}} = \left ( {{32} \over {27}} \right ) \; {{{t^{2/3}} \, {\sin ^4} \theta } \over {F^3}} \; \; \; .
\eqno(3.13)
$$
The metric is type I.

We now use the same seed metric but with the Killing vector 
$ {\xi^a} = {{(\partial / \partial \phi )}^a} $.
Then we have
$ \lambda = {t^{2/3}} \, {r^2} $
and
$ \omega = 0 $.
It then follows that
$ F = {\cos ^2} \theta \; + \; {t^{4/3}} \, {r^4} \,  {\sin ^2} \theta \,  $.
The equation for 
$ \alpha _a $
is then
$$
{\nabla _{[a}} {\alpha _{b]}} =  {2 \over 3} \; r \, {\nabla _{[a}} r \, {\nabla _{b]}} z \; + \; 2 \, {t^{1/3}} \, {\nabla _{[a}} t \, {\nabla _{b]}} z  
\eqno(3.14)
$$
(as well as $ {\xi ^a} {\alpha _a} = 0 $).
A solution is 
$$
{\alpha _a} = \left ( {1 \over 3} \; {r^2} \; + \; {3 \over 2} \; {t^{4/3}} \right ) \; {\nabla _a} z \; \; \; .
\eqno(3.15)
$$
The one-form
$ \eta _a $
is then given by
$$
{\eta _a} = {\nabla _a} \phi \; + \; \left ( {2 \over 3} \; {r^2} \; + \; 3 \; {t^{4/3}} \right )  \; \cos \theta \, \sin \theta  \, {\nabla _a} z \; \; \; .
\eqno(3.16)
$$
The new metric 
$ {\tilde g}_{ab} $
is then given by
$$
d {{\tilde s}^2} = F \; \left [ - \; d {t^2} \; + \; {t^{2/3}} \; \left ( d {z^2} \; + \; d {r^2} \right ) \right ] 
$$
$$
+ \; {{{t^{2/3}} \, {r^2}} \over F}  \; {{\left ( d \phi \; + \; \left [ {2 \over 3} \; {r^2} \; + \; 3 \; {t^{4/3}} \right ] \; \cos \theta \, \sin \theta  \, d z \right ) }^2} \; \; \; .
\eqno(3.17)
$$
This metric has no singularities on the axis.  (There is a conical singularity
if the range of $ \phi $ is $ 2 \pi $.  However, this singularity can be
removed by choosing the appropriate range of $ \phi $).
The vorticity vanishes.  The expansion and acceleration are given by
$$
{\tilde \Theta} = {1 \over {\sqrt F}} \; \left ( {t^{- 1}} \; + \; {2 \over 3} \; {F^{ - 1}} {t^{1/3}} {r^4} {\sin ^2} \theta \right ) \; \; \; ,
\eqno(3.18)
$$
$$
{{\tilde a}_a} = 2 {F^{ - 1}} {t^{4/3}} {r^3} {\sin ^2} \theta \, {\nabla _a} r 
\; \; \; .
\eqno(3.19)
$$
The shear has components in the $ z z , \, r r , \, \phi \phi $
and $ z \phi $ directions.  Its square is
$$
{{\tilde \sigma}^{ab}} {{\tilde \sigma }_{ab}} = \left ( {8 \over {27}} \right ) \; {{{r^2} \, {t^{2/3}} \, {\sin ^2} \theta } \over {F^3}} \; \left [ 27 \, {\cos ^2} \theta \, + \, 4 {r^6} {\sin ^2} \theta \right ] \; \; \; .
\eqno(3.20)
$$
The metric is type I.

\subsection{Tabensky-Taub seed metric}
 
We now consider a different seed metric: The Tabensky-Taub metric\cite{taub}
$$
ds^2=-Vdt^2+Vdz^2+z(dx^2+dy^2) 
\eqno(3.21)
$$
where
$$
V = z^{-1/2}\exp (\frac 12a^2z^2) \; \; \; .
\eqno(3.22)
$$
This is a solution of the Einstein-perfect-fluid equations with
$$
P = \rho = {{a^2} \over {16 \pi V}} \; \; \; .
\eqno(3.23)
$$
The fluid four-velocity is
$ {u_a} = - \, {\sqrt V} \, {\nabla _a} t $.
The expansion, vorticity and shear vanish.  The acceleration is 
$$
{a_a} = {1 \over {2 V}} \; {{d V} \over {d z}} \; {\nabla _a} z \; \; \; .
\eqno(3.24)
$$
The metric is type D.

First we consider the Killing vector
$ {\xi ^a} = {{( \partial / \partial x )}^a} $.
Then we have
$ \lambda = z $
and thus
$ F = \cos ^2\theta +z^2\sin ^2\theta $.
The equation for 
$ \alpha _a $
is
$ {\nabla _{[a}} {\alpha _{b]}} =  - \;  {\nabla _{[a}} y \, {\nabla _{b]}} t $
(as well as $ {\xi ^a} {\alpha _a} = 0 $).
The solution is 
$ {\alpha _a} = - \, y \, {\nabla _a} t $.
It then follows that
$$
{\eta _a} = {\nabla _a} x \; - \; 2 \, \cos \theta \, \sin \theta \, y \, {\nabla _a} t \; \; \; .
\eqno(3.25)
$$
The new metric is then given by
$$
d {{\tilde s}^2} = F \, \left ( - \, V d {t^2} \; + \; V d {z^2} \; + \; z d {y^2} \right ) \; + \; {z \over F} \; {{\left ( d x \; - \;  2 \, \cos \theta \, \sin \theta \, y \, d t \right ) }^2} \; \; \; .
\eqno(3.26)
$$
Both the seed metric and the new metric have singularities at $ z = 0 $.
However, the new solution does not appear to have any additional singularities.
The expansion and the vorticity vanish.  The acceleration is 
$$
{{\tilde a}_a} = \left ( {1 \over {2 V}} \; {{d V} \over {d z}} \; + \; {{z {\sin ^2} \theta } \over F} \right ) \, {\nabla _a} z \; \; \; .
\eqno(3.27)
$$
The shear has components in the $x y $ and $ t y $ directions.  Its square is
$$
{{\tilde \sigma }^{ab}} {{\tilde \sigma }_{ab}} = {{2 \, {\cos ^2} \theta \, {\sin ^2} \theta } \over {{F^3} \, V}} \; \; \; .
\eqno(3.28)
$$
The metric is type I.

We now consider another Killing vector of the Tabensky-Taub metric.  Using
polar coordinates for the $ x-y $ plane we find that the seed metric is
$$
d {s^2} = -Vdt^2+Vdz^2+z(dr^2+ {r^2} d {\phi ^2}) \; \; \; .
\eqno(3.29)
$$
Using the Killing vector
$ {\xi ^a} = {{(\partial / \partial \phi )}^a} $
we find
$ \lambda = z \, {r^2} $
and
$ \omega = 0 $.
Therefore
$ F =  {\cos ^2} \theta \; + \; {z^2} \, {r^4} \,  {\sin ^2} \theta \, $.
The equation for 
$ \alpha _a $
is 
$$
{\nabla _{[a}} {\alpha _{b]}} = r \, {\nabla _{[a}} r \, {\nabla _{b]}} t \; + \; 2 \, V \, {\nabla _{[a}} t \, {\nabla _{b]}} z
\eqno(3.30)
$$
(as well as $ {\xi ^a} {\alpha _a} = 0 $).
A solution is
$$
{\alpha _a} = 2 \, V \, t \, {\nabla _a} z \; - \; r t \, {\nabla _a} r \; \; \; .
\eqno(3.31)
$$
The vector
$ \eta _a $
is then given by
$$
{\eta _a} = {\nabla _a} \phi \; + \; 2 \cos \theta \, \sin \theta \, \left (   2 \, V \, t \, {\nabla _a} z \; - \; r t \, {\nabla _a} r \right ) \; \; \; .
\eqno(3.32)
$$
The new metric is then given by
$$
d {{\tilde s}^2} = F \left ( -Vdt^2+Vdz^2+z dr^2 \right ) 
$$
$$
+ \; {{z {r^2}} \over F} \; {{\left ( d \phi \; + \; 2 \cos \theta \, \sin \theta \, \left [ 2 V t \, d z \, - \, r t \, d r \right ] \right ) }^2} \; \; \; .
\eqno(3.33)
$$
This metric does not have any singularity on the axis (though one must choose the range of $ \phi $ to avoid a conical singularity).  The expansion and vorticity vanish.  The acceleration is
$$
{{\tilde a}_a} = \left ( {1 \over {2 V}} \; {{d V} \over {d z}} \; + \; {{z {r^4} {\sin ^2} \theta } \over F} \right ) \, {\nabla _a} z \; + \; {{2 {z^2} {r^3} {\sin ^2} \theta } \over F} \; {\nabla _a} r \; \; \; .
\eqno(3.34)
$$
The shear has components in the
$ r r , \, r \phi , \, r z , \, z z $
and $ \phi z $
directions.  Its square is
$$
{{\tilde \sigma }^{ab}} {{\tilde \sigma }_{ab}} = {{2 \, {r^2} \, {\cos ^2} \theta \, {\sin ^2}
\theta \, \left ( {r^2} + 4 z V \right ) } \over {{F^3} \, V}} \; \; \; .
\eqno(3.35)
$$
The metric is type I.

\subsection{Allnutt seed metric}

Our next seed metric is due to Allnutt.\cite{allnutt}  It is
$$
d {s^2} = {z^2} \; \left [ - \; {{d {t^2}} \over {1 + {t^2}}} \; + \; f \, d {x^2} \; + \; {{{t^2} \, \left ( 1 + {t^2} \right ) } \over f} \; d {y^2} \right ] \; + \; d {z^2} 
\eqno(3.36)
$$
where the function 
$ f $
is 
$ f = {t^{ 2 \beta }} \, {{\left ( 1 + {t^2} \right ) }^{1 - \beta }} $
and
$ \beta $
is a constant.  This spacetime is a solution of the Einstein-perfect fluid
equations with
$$
P = \rho = {{\beta \, ( 1 - \beta )} \over {8 \pi \, {z^2} \, {t^2} \, \left ( 1 + {t^2} \right ) }} 
\eqno(3.37)
$$
and 
$ {u_a} = - \, z \, {{\left ( 1 + {t^2} \right ) }^{- 1 /2}} {\nabla _a} t $.
The vorticity vanishes.  The acceleration is 
$ {a_a} = {z^{ - 1}} {\nabla _a} z $
and the expansion is
$$
\Theta = {{2 {t^2} + 1} \over {z t {\sqrt {1 + {t^2}}}}} \; \; \; .
\eqno(3.38)
$$
The shear has components in the
$ xx , \, yy $
and
$ zz $
directions.  Its square is
$$
{\sigma ^{ab}} {\sigma _{ab}} = {2 \over {3 {z^2} {t^2} \left ( 1 + {t^2} \right ) }} \; \left [ {t^4} \, + \,  {t^2} \, + \, 1 \, + \, 3 \beta ( \beta - 1 ) \right ] \; \; \; .
\eqno(3.39)
$$
The metric is type D.
  
Using the Killing vector
$ {\xi ^a} = {{(\partial / \partial x )}^a} $
we find
$ \lambda = {z^2} f $
and thus
$ F = {\cos ^2} \theta \, + \, {z^4} {f^2} \, {\sin ^2} \theta $.
The equation for
$ \alpha _a $
is 
$$
{\nabla _{[a}} {\alpha _{b]}} = 2 \, {z^2} \, t \, {\nabla _{[a}} t \, {\nabla _{b]}} y \; + \; 2 \, z \, \left ( {t^2} + \beta \right ) \, {\nabla _{[a}} z \, {\nabla _{b]}} y 
\eqno(3.40)
$$
(as well as $ {\xi ^a} {\alpha _a} = 0 $).
A solution is 
$ {\alpha _a} = {z^2} \, \left ( {t^2} + \beta \right ) \, {\nabla _a} y $.
It then follows that
$$
{\eta _a} = {\nabla _a} x \; + \; 2 \cos \theta \sin \theta \, {z^2} \left ( {t^2} + \beta \right ) \, {\nabla _a} y \; \; \; .
\eqno(3.41)
$$
The new metric is then given by
$$
d {{\tilde s}^2} = F \; \left [ - \; {{z^2} \over {1 + {t^2}}} \; d {t^2} \; + \; {{{z^2} {t^2} \left ( 1 + {t^2} \right ) } \over f} \; d {y^2} \; + \; d {z^2} \right ] 
$$
$$
+ \; {{{z^2} f} \over F} \; {{\left [ d x \; + \; 2 \cos \theta \sin \theta \, {z^2} \, \left ( {t^2} + \beta \right ) \, d y \right ] }^2} \; \; \; .
\eqno(3.42)
$$
Both the seed metric and the new metric are singular at $ z = 0 $
and at $ t = 0 $.  However, the new metric does not appear to have any
additional singularities.  The vorticity vanishes.  The expansion is
$$
{\tilde \Theta } = {\sqrt {{1 + {t^2}} \over F}} \; \left ( {{2 {t^2} + 1} \over {z t \left ( 1 + {t^2} \right ) }} \; + \; {{{z^3} f {\dot f} {\sin ^2} \theta } \over F} \right )
\eqno(3.43)
$$
where an overdot denotes derivative with respect to $ t $.  The acceleration is
$$
{{\tilde a}_a} = \left ( {z^{ - 1}} \; + \; 2 {F^{ - 1}} {z^3} {f^2} {\sin ^2} \theta \right ) \, {\nabla _a} z \; \; \; .
\eqno(3.44)
$$
The shear has components in the $x x , \, y y , \, z z $
and $ x y$ directions.  Its square is
$$
{{\tilde \sigma}^{ab}} {{\tilde \sigma }_{ab}} = {2 \over {3 {z^2} {t^2} ( 1 + {t^2} ) F}} \; \left [ {t^4} \, + \, {t^2} \, + \, 1 \, + \, 3 \beta ( \beta - 1) \right ]
$$
$$
+ \; {8 \over 3} \; {\sin ^2} \theta \, {z^2} \; {{f^2} \over {F^2}} \; \left [ {{{t^2} + \beta } \over t} \; \left ( {{\dot F} \over F} \; + \; {{1 - 3 \beta - {t^2}} \over {t ( 1 + {t^2} )}} \right ) \; + \; {{3 {\cos ^2} \theta } \over F} \right ] \; \; \; .
\eqno(3.45)
$$
The metric is type I.

\subsection{Kramer seed metric}

The final $ P = \rho $ seed metric that we shall consider is due to Kramer.\cite{kramer}  It is
$$
d {s^2} = {e^{{\alpha ^2} {r^2}}} \; \left ( - \, d {t^2} \, + \, d {r^2} \right ) \; + \; {r^2} \, d {\phi ^2} \; + \; d {z^2} 
\eqno(3.46)
$$
where $ \alpha $ is a constant.  This is a solution of the Einstein-perfect
fluid equations with 
$$
P = \rho = {{\alpha ^2} \over {8 \pi}} \; {e^{ - \, {\alpha ^2} {r^2}}} \; \; \; .
\eqno(3.47)
$$
The four velocity is 
$ {u_a} = - \, {e^{{\alpha ^2} {r^2}/2}} \, {\nabla _a} t $.
The expansion, shear and vorticity vanish.  The acceleration is
$ {a_a} = {\alpha ^2} r {\nabla _a} r $. 
The metric is type D. 
We use the Killing vector
$ {\xi ^a} = {{( \partial / \partial \phi )}^a} $.
Then we have
$ \lambda = {r^2} $
and thus
$ F = {\cos ^2} \theta \, + \, {r^4} {\sin ^2} \theta $.
The equation for 
$ \alpha _a $
is 
$ {\nabla _{[a}} {\alpha _{b]}} = - \, 2 \, {\nabla _{[a}} z \, {\nabla _{b]}} t $
(as well as ${\xi ^a} {\alpha _a} = 0 $).
A solution is
$ {\alpha _a} = - \, 2 z {\nabla _a} t $.
The new metric is then given by
$$
d {{\tilde s}^2} = F \, \left [ {e^{{\alpha ^2} {r^2}}} \, \left ( - \, d {t^2} \, + \, d {r^2} \right ) \; + \; d {z^2} \right ] \; + \; {{r^2} \over F} \; {{\left ( d \phi \, - \, 4 \cos \theta \sin \theta \, z \, d t \right ) }^2}
\; \; \; .
\eqno(3.48)
$$
The new metric has no singularities on the axis, though one must choose the 
range of $ \phi $ to avoid a conical singularity.  The expansion and vorticity
vanish.  The acceleration is
$$
{{\tilde a}_a} = \left ( {\alpha ^2} r \; + \; 2 {F^{ - 1}} {r^3} {\sin ^2} \theta \right ) {\nabla _a} r \; \; \; .
\eqno(3.49)
$$
The shear has components in the $z \phi $ and $ z t$ directions.  Its square is
$$
{{\tilde \sigma}^{ab}} {{\tilde \sigma }_{ab}} = {{8 \, {r^2} \, {e^{ - \, {\alpha ^2} {r^2}}} \, {\cos ^2} \theta {\sin ^2} \theta } \over {F^3}} \; \; \; .
\eqno(3.50)
$$
The metric is type I.

\section{$ \rho + 3 P = 0 $ fluids}

\subsection{Fluid variables}

We now consider the solutions with the equation of state 
$\rm \rho + 3P = 0$.  This equation of state occurs in the Einstein 
static universe\cite{he}, in one of the limiting solutions of the Wahlquist 
metric\cite{wahlquist} due to Vaidya\cite{exact,vaidya} and has been 
suggested as the equation 
of state for a chaotic array of strings\cite{strings}.
In generating solutions with the equation of state $ \rho + 3 P = 0 $
we must choose the fluid four-velocity of the seed metric to be parallel 
to the Killing vector.  The fluid four velocity of the new metric will then
also be parallel to the Killing vector.  This results in considerable 
simplification of the fluid parameters.  The expansion and shear of both
the seed solution and the new solution vanish.  The density, four-velocity 
and acceleration transform as follows:
$$
{\tilde \rho} = \rho / F \; \; \; ,
\eqno(4.1)
$$
$$
{{\tilde u}^a} = {\sqrt F} \, {u^a} \; \; \; ,
\eqno(4.2)
$$
$$
{{\tilde a}_a} = \left ( 2 {F^{ - 1}} {\cos ^2} \theta \, - \, 1 \right ) \, {a_a} \; \; \; .
\eqno(4.3)
$$
Since we choose seed solutions with zero twist, it follows that the fluid 
vorticity of the seed solution vanishes.  The vorticity of the new metric is
$$
{{\tilde \omega}_{ab}} = 2 \cos \theta \sin \theta \, {\sqrt {- \, \lambda /F}} \; {\nabla _{[a}} {\alpha _{b]}} \; \; \; .
\eqno(4.4)
$$
It then follows that the vorticity vector 
$ {{\tilde \Omega }_a} := {{\tilde \epsilon }_{abcd}} {{\tilde u}^b} {{\tilde \omega }^{cd}} $
is parallel to the acceleration vector $ {\tilde a}_a $.

\subsection{Vaidya seed metric}

Our first seed metric is due to Vaidya\cite{vaidya} and is given by
$$
d {s^2} = - \, f \, d {t^2} \; + \; {{d {r^2}} \over {\left ( 1 + c {r^2} \right ) \, f}} \; + \; {r^2} \; \left ( d {\psi ^2} \; + \; {\sin ^2} \psi d {\phi ^2} \right ) \; \; \; .
\eqno(4.5)
$$
Here
$$
f := 1 \; - \; {{2 m} \over r} \; {\sqrt {1 + c {r^2}}} 
\eqno(4.6)
$$
and $ c $ and $ m $ are constants.  This two parameter family of solutions
contains the Schwarzschild metric and the Einstein static universe as special 
cases.  The energy density is
$ \rho = - \, 3 c f /(8 \pi ) $
and the fluid four velocity is 
$ {u_a} = - \, {\sqrt f} \, {\nabla _a} t $.
The acceleration is 
$$
{a_a} = {1 \over {2 f}} \; {{d f} \over {d r}} \; {\nabla _a} r \; \; \; .
\eqno(4.7)
$$
The metric is type D.
The timelike Killing vector is 
$ {\xi ^a} = {{(\partial / \partial t)}^a} $.
It then follows that 
$ \lambda = - \, f $
and therefore that
$ F = {\cos ^2} \theta \, + \, {f^2} {\sin ^2} \theta $.
The equation for 
$ \alpha _a $
is
$ {\nabla _{[a}} {\alpha _{b]}} = - \, 2 m \, \sin \psi \, {\nabla _{[a}} \psi \, {\nabla _{b]}} \phi $
(as well as $ {\xi ^a} {\alpha _a} = 0 $).  A solution is
$ {\alpha _a} = 2 m \, \cos \psi \, {\nabla _a} \phi $.
We then find
$$
{\eta _a} = {\nabla _a} t \; + \; 4 m \, \cos \theta \sin \theta \, \cos \psi \, {\nabla _a} \phi \; \; \; .
\eqno(4.8)
$$
The new metric is 
$$
d {{\tilde s}^2} = F \, \left [ {{d {r^2}} \over {\left ( 1 + c {r^2} \right ) \, f}} \; + \; {r^2} \, \left ( d {\psi ^2} \, +  \, {\sin ^2} \psi \, d {\phi ^2} \right ) \right ] \; - \; {f \over F} \; {{\left ( d t \, + \, 4 m \cos \theta \sin \theta \cos \psi d \phi \right ) }^2} \; \; \; .
\eqno(4.9)
$$
This metric has a singularity on the axis.  The acceleration is
$$
{{\tilde a}_a} = \left ( 2 {F^{ - 1}} \, {\cos ^2} \theta \, - \; 1 \right ) \; {1 \over {2 f}} \; {{d f} \over {d r}} \; {\nabla _a} r \; \; \; .
\eqno(4.10)
$$
The nonvanishing components of the vorticity are
$$
{{\tilde \omega }_{\psi \phi}} = - \, {{\tilde \omega}_{\phi \psi}} = - \, 2 m \cos \theta \sin \theta \; {\sqrt {f \over F}} \; \sin \psi \; \; \; .
\eqno(4.11)
$$
After the transformation the pressure and energy density are 
scaled by $ F^{-1}$ from their static values.  As in the static case, 
on the horizon the pressure and scalar curvature are zero.  The 
generated vorticity tensor vanishes on the horizon.  The metric is type I. 

\subsection{Kramer seed metric}

The next seed metric that we will consider is essentially due to Kramer.\cite{kramer}  It is
$$
d {s^2} = - \, {e^{2 x}} \, d {t^2} \; + \; {1 \over {{e^{2x}} \, \sinh b x}} \; \left [ {e^{{\sqrt {{b^2} - 4}} \, x}} \, d {\phi ^2} \; + \; {e^{ - \, {\sqrt {{b^2} - 4}} \, x}} \, d {z^2} \; + \; {{d {x^2}} \over {\sinh b x}} \right ] \; \; \; .
\eqno(4.12)
$$
Here $ b $ is a constant.  The energy density is
$ \rho = 3 {b^2} {e^{2x}}/(32 \pi )$
and the fluid four velocity is
$ {u_a} = - \, {e^x} {\nabla _a} t $.
The acceleration is 
$ {a_a} = {\nabla _a} x $.
The metric is type I.
The Killing vector is
$ {\xi ^a} = {{(\partial / \partial t)}^a} $
and therefore we have
$ \lambda = - \, {e^{2x}} $.
It then follows that
$ F = {\cos ^2} \theta \, + \, {e^{4 x}} \, {\sin ^2} \theta $.
The equation for 
$ \alpha _a $
is
$ {\nabla _{[a}} {\alpha _{b]}} = 2 \, {\nabla _{[a}} z \, {\nabla _{b]}} \phi $ (as well as $ {\xi ^a} {\alpha _a} = 0 $).  A solution is
$ {\alpha _a} = 2 z {\nabla _a} \phi $.
We then find
$$
{\eta _a} = {\nabla _a} t \, + \, 4 \cos \theta \sin \theta z \, {\nabla _a} \phi \; \; \; .
\eqno(4.13)
$$
The new metric is then given by
$$
d {{\tilde s}^2} = {F \over {{e^{2x}} \, \sinh b x}} \; \left [   {e^{{\sqrt {{b^2} - 4}} \, x}} \, d {\phi ^2} \; + \; {e^{ - \, {\sqrt {{b^2} - 4}} \, x}} \, d {z^2} \; + \; {{d {x^2}} \over {\sinh b x}} \right ]  \; - \; {{e^{2 x}} \over F} \; {{\left ( d t \, + \, 4 \cos \theta \sin \theta \, z \, d \phi \right ) }^2} \; \; \; .
\eqno(4.14)
$$
The solution generating process does not seem to have added any new singularities.  The acceleration is
$$
{{\tilde a}_a} = \left ( 2 {F^{ - 1}} {\cos ^2} \theta \, - \, 1 \right ) \, {\nabla _a} x \; \; \; .
\eqno(4.15)
$$
The nonvanishing components of the vorticity are
$$
{{\tilde \omega}_{z \phi}} = - \, {{\tilde \omega}_{\phi z}} = 2 \cos \theta \sin \theta {e^x} {F^{ - 1}} \; \; \; .
\eqno(4.16)
$$
The metric is type I.

\section{Discussion}

Exact solutions to the Einstein field equations are valuable as a tool 
in the exploration of behavior ranges allowed by the field equations.  
With the marked increase in numerical calculations, exact solutions are 
also useful as comparisons for approximate or numerical solutions and as 
checks for the development of computer codes.  In general Einstein's 
equations for static spaces are far easier to solve than those for 
stationary spacetimes.  One advantage of solution generating methods is 
that they can be used to find stationary solutions using simple static 
seed solutions.  In this paper, we have used the method of Stephani\cite{stephani} 
to generate stationary solutions of the Einstein-perfect fluid field 
equation from simple static seeds.  A possible disadvantage of solution 
generating is that often extra singularities, especially on an axis, can 
be introduced.  Some of the solutions that we generated have no new 
singularities and in some cases they appear to be non-singular.

While solution generating techniques have been applied extensively to 
vacuum spacetimes and to solutions of the Einstein-Maxwell equations\cite{exact}; with some notable exceptions\cite{taub,wainwright}
there is comparatively little work on solution generating involving
perfect fluids.  The method of reference\cite{wainwright} generates
$P=\rho$ perfect fluid solutions from vacuum solutions.  It can be applied
only to solutions with two spacelike Killing vectors and generates only solutions with zero vorticity.  In contrast the Stephani method requires
only one Killing vector and is compatible with nonzero vorticity.

There has been recent numerical work on vacuum cosmologies with two 
spacelike Killing vectors.\cite{beverly}  In this work the known
vacuum exact solutions provided useful test cases.  The numerical 
work of reference\cite{beverly} could easily be generalized to the 
case of inhomogeneous perfect fluid cosmologies.  The exact solutions
that we have generated in this paper could be used as test cases for
such numerical work.
  
All of  the seed metrics that we have used 
belong to the Perjes Class II\cite{perjes}.  
Since this classification is preserved under the transformation\cite{jean} the 
generated metrics are also members of this class. For the 
$\rm P = - \, \rho /3$ solutions, the generated 
solutions are all shear and expansion 
free and all have a non-zero vorticity that lies along the static 
acceleration.  This is a characteristic of stationary Perjes II 
solutions.  No assumption about algebraic speciality is made.  In the 
$\rm P = \rho$ case, the fluid parameters of the new solutions are not 
constrained to be shear free or rigidly rotating.  The action of the 
transformation in this case is on the spacelike Killing vector and all 
of the generated solutions as well as their seeds have zero vorticity 
while their shear, expansion and acceleration structure can be quite 
complex. Solution generating allows access to exact solutions that are 
difficult to find using other methods.  These new solutions illustrate
various behaviors allowed by the Einstein-perfect fluid equations 
and may additionally serve as aids in 
further numerical work.  In some cases they provide insights into the 
static origins of existing solutions obtained through more arduous 
methods.
 
\section{Acknowledgements}

We thank the University of Michigan for hospitality.  D.G. was partially 
supported by NSF Grant PHY9408439 to Oakland University and 
by a Cottrell College Science Award of Research Corporation 
to Oakland University. 
E.N. Glass was partially supported by an NSERC of Canada grant.
Computations were verified using MapleV.3 (Waterloo Maple Software, Waterloo,
Ontario) and GRTensorII (P. Musgrave, D.Pollney, and K. Lake, Queens
University, Kingston, Ontario).


\begin{references}

\bibitem{ehlers}
J. Ehlers, {\it Les theories relativistes de la gravitation} (CNRS, Paris,
1959)

\bibitem{harrison}
B. K. Harrison, J. Math. Phys. {\bf 9}, 1744 (1968)

\bibitem{geroch}
R. Geroch, J. Math. Phys. {\bf 12}, 918 (1971)

\bibitem{kramer1}
D. Kramer and G. Neugebauer, Commun. Math. Phys. {\bf 10}, 132 (1968)

\bibitem{hauern}
I. Hauser and F. Ernst, J. Math. Phys. {\bf 19}, 1316 (1978)

\bibitem{kramer2}
G. Neugebauer and D. Kramer, Ann. Physik {\bf 24} , 62 (1969) 
\bibitem{stephani}
H. Stephani, J. Math. Phys. {\bf 29}, 1650 (1988)

\bibitem{exact} {\it Exact Solutions of Einstein's Field Equations} by
D. Kramer, H. Stephani, E. Herlt, M. MacCallum and E. Schmutzer, Cambridge
University Press (Cambridge, U.K. 1980)

\bibitem{taub}
R. Tabensky and A.H. Taub, Commun. Math. Phys. {\bf 29}, 61 (1973)

\bibitem{wainwright}
J. Wainwright, W.C.W. Ince and B.J. Marshman, Gen. Rel. Grav. {\bf 10}, 259 (1979)

\bibitem{allnutt}
J.A. Allnutt, Gen. Rel. Grav. {\bf 13}, 1017 (1981)

\bibitem{kramer}
Kramer, Class. Quantum Grav. {\bf 5}, 393 (1988)

\bibitem{he}
{\it The Large Scale Structure of Space-Time} by S.W. Hawking and G.F.R. Ellis,
Cambridge University Press (Cambridge, U.K. 1973)

\bibitem{wahlquist}
H.D. Wahlquist, Phys. Rev. {\bf 172}, 1291 (1968)

\bibitem{vaidya}
P.C. Vaidya, Pramana {\bf 8}, 512 (1977)

\bibitem{strings} A. Vilenkin, Phys. Rev. D {\bf 24}, 2082 (1981)

\bibitem{beverly}
B. Berger and V. Moncrief, Phys. Rev. D {\bf 48}, 4676 (1993); 
B. Berger, D. Garfinkle and V. Swamy, Gen. Rel. Grav. {\bf 27}, 511 (1995)

\bibitem{perjes}
Z. Perjes, Int. J. Theor. Phys. {\bf 10}, 217 (1974); B. Lukacs, Z. Perjes 
and A. Sebesstyen, Gen. Rel. Grav. {\bf 15}, 511 (1983).

\bibitem{jean}
J.P. Krisch, J. Math. Phys. {\bf 29}, 447 (1988).

\end{references}
\end{document}